\documentclass[12pt]{article}
\usepackage{epsfig,pstricks}

\newcommand{\nc}{\newcommand}
\nc{\beq}{\begin{equation}}
\nc{\eeq}{\end{equation}}
\nc{\beqa}{\begin{eqnarray}}
\nc{\eeqa}{\end{eqnarray}}

\newwrite\ffile\global\newcount\figno \global\figno=1

\def\writedef#1{}

\input epsf
\def\figin{\epsfcheck\figin}\def\figins{\epsfcheck\figins}
\def\epsfcheck{\ifx\epsfbox\UnDeFiNeD
\message{(NO epsf.tex, FIGURES WILL BE IGNORED)}
\gdef\figin##1{\vskip2in}\gdef\figins##1{\hskip.5in}
\else\message{(FIGURES WILL BE INCLUDED)}%
\gdef\figin##1{##1}\gdef\figins##1{##1}\fi}

\def\figinsert{}
\def\ifig#1#2#3{\xdef#1{fig.~\the\figno}
\writedef{#1\leftbracket fig.\noexpand~\the\figno}%
\figinsert\figin{\centerline{#3}}\medskip\centerline{\vbox{\baselineskip12pt
\advance\hsize by -1truein\center\footnotesize{  Fig.~\the\figno.} #2}}
\bigskip\endinsert\global\advance\figno by1}
\def\endinsert{}

\begin{document}

\title{\large{\bf Triviality and the Precision Bound on the Higgs Mass}}

\author{
R. Sekhar Chivukula\thanks{sekhar@bu.edu} \\
{\small\em Department of Physics,
Boston University, Boston, MA 02215, USA.} \\ 
Nick Evans\thanks{n.evans@phys.soton.ac.uk}
\\ {\small\em Department of Physics,
University of Southampton, Southampton, S017 1BJ, UK.} }

\date{July 1999}

\maketitle

\begin{picture}(0,0)(0,0)
\put(350,275){BUHEP-99-15}
\end{picture}

\begin{abstract}

  The triviality of the scalar sector of the standard one-doublet Higgs
  model implies that this model is only an effective low-energy theory
  valid below some cut-off scale $\Lambda$. For a heavy higgs this scale
  must be relatively low (10 TeV or less).  Additional interactions
  coming from the underlying theory, and suppressed by the scale
  $\Lambda$, give rise to model-dependent corrections to precisely
  measured electroweak quantities. Dimension six operators arising from
  the underlying physics naturally contribute to the $S$ and $T$
  parameters, and their effects should be included in a global fit to
  the precision data that determines any limit on the Higgs mass. Using
  dimensional analysis, we estimate the expected size of these
  corrections in a custodially-symmetric strongly-interacting underlying
  theory .  Taking these operators' coefficients to be of natural size
  gives sufficiently large contributions to the T parameter to reconcile
  Higgs masses as large as 400-500 GeV with the precision data.

\end{abstract}

\newpage

The standard one-doublet Higgs model is, at first sight, a fully
consistent, renormalizable, quantum field theory. Order by order in
perturbation theory, all experimentally measurable quantities are
completely calculable (see, for example, ref. \cite{Erler:1998ig} and
references therein) in terms of the gauge coupling constants, the weak
scale $v\approx 250$ GeV, the quark and lepton masses and mixing angles
(most importantly, the top-quark mass), and the Higgs boson mass.
Conversely, using sufficiently many measurements one may test the
consistency of the data with the standard Higgs model, and extract the
best-fit values of the parameters used to define the theory. This method
results in an indirect determination of the Higgs boson mass which is
relatively low, $m_H < 262$ GeV (95\% C.L.) \cite{Karlen}. This bound is
only relevant for the standard one-doublet Higgs model in the absence of
new physics at higher energies
\cite{newphysics}.

However, the triviality \cite{Wilson}
of the scalar sector of the standard one-doublet Higgs model implies
that the model is only an effective low-energy theory valid below some
cut-off scale $\Lambda$. Additional interactions coming from the
underlying theory\footnote{Examples of such theories include
  top-condensate
  \protect\cite{topcondensate,seesaw,composite}
  and composite Higgs models
  \protect\cite{composite}.}, and
suppressed by the scale $\Lambda$, give rise to model-dependent
corrections to precisely measured electroweak quantities. In this sense
the standard model, for a given Higgs boson mass, is not a single theory
but rather a class of theories.  The most important corrections from the
underlying theory are encoded in dimension six operators
\cite{effective}
which contribute to the Peskin-Takeuchi $S$ and $T$ parameters
\cite{Peskin}. Their contribution should be
included in a global fit to the precision data that determines a limit
on the Higgs mass \cite{newphysics}.
Here we emphasize the role that triviality bounds have in this context,
compare the precision bounds with the triviality bound, and discuss the
natural size of the couplings and the scale $\Lambda$ for a
strongly-interacting underlying theory.

Clearly, if $\Lambda$ can be taken to be arbitrarily high, the
corrections from the underlying theory will be irrelevant.  Because of
triviality, however, for a given Higgs boson mass the scale $\Lambda$
cannot be arbitrarily high \cite{Dashen:1983ts}.  An estimate of the
upper bound on the cut-off can be taken from lowest order perturbation
theory. Integrating the lowest order beta function for the Higgs self
coupling $\lambda$
\beq
\beta(\lambda) = \mu {d \lambda \over d \mu} = {3 \over 2 \pi^2} \lambda^2 
+ ...
\eeq
one finds
\beq
{1 \over \lambda(\mu)} - {1 \over \lambda(\Lambda)} =  {3 \over 2 \pi^2} 
\log {\Lambda \over \mu}~.
\eeq
Using the relation $m_H^2 = 2 \lambda(m_H) v^2$ then implies
\beq
m_H^2 \log {\Lambda \over m_H} \leq {4 \pi^2 v^2 \over 3}
\label{bound}
\eeq

\begin{figure}
\epsfxsize 10cm \centerline
{\epsffile{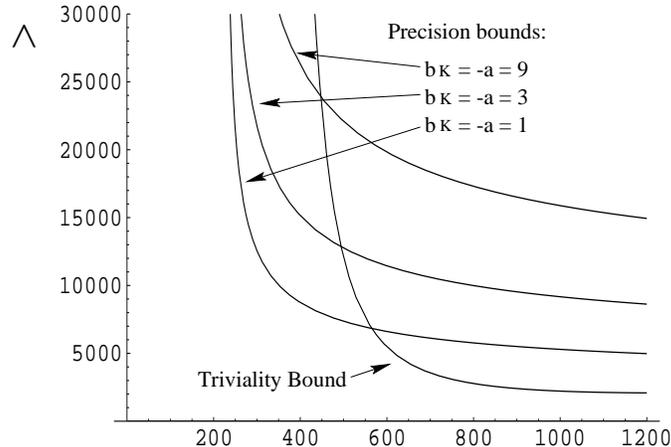}}
\caption{
The triviality and precision bounds in the $\Lambda$-$m_h$ plane. The regions
above the corresponding curves are excluded.
}
\label{Fig1}
\end{figure}

The resulting upper bound on $\Lambda$ as a function of $m_H$ is shown
in Figure \ref{Fig1}. If the underlying theory comes close to saturating
the upper bound, then $\lambda(\Lambda)$ is large and the
underlying theory must be strongly interacting. A theory with a Higgs
mass above 500 GeV, for example, must have a cut off below about 12 TeV.
This derivation is based on perturbation theory and would appear
suspect, but non-perturbative
\cite{lattice}
studies on the lattice using analytic and Monte Carlo techniques confirm
that the estimate in eqn. (\ref{bound}) is reasonably accurate.

To estimate the sizes of effects from the underlying physics, we will
rely on dimensional analysis \cite{Georgi:1993dw}.  In brief, a theory
with light scalar particles depends on two parameters: $\Lambda$, the
energy scale of the underlying physics, and $\kappa$, a measure of the
size of dimensionless couplings (in the chiral lagrangian in QCD $\kappa
= {\cal O}(\Lambda_ {\chi SB}/ f_\pi$)
\cite{powercounting}).  For a strongly-coupled
underlying theory, $\kappa$ is expected of order $4\pi$.
Starting from the kinetic energy term (which is bilinear in the scalar
field), the sizes of operators in the effective low-energy theory can be
estimated by including an extra value of $\kappa$ for each scalar field
(beyond the two present in the kinetic energy) and making up the
mass-dimension by using the appropriate power of $\Lambda$.  The
leading operators in the effective lagrangian
\cite{effective}
which contribute to electroweak measurements are then, 
\beq
-\,{a \over 2!\,\Lambda^2} \left\{ [D_\mu,D_\nu] \phi \right\}^\dagger[D^\mu,D^\nu]
\phi
+{\tilde{b}\, \kappa^2 \over 2!\, \Lambda^2} (\phi^\dagger \stackrel{\leftrightarrow}{D^\mu} \phi) 
(\phi^\dagger \stackrel{\leftrightarrow}{D_\mu} \phi)~,
\label{operators}
\eeq
where $a$ and $\tilde{b}$ are expected to be of order one
\cite{Chivukula:1996sn,Chivukula:1997iw}.

The second term in eqn. (\ref{operators}) violates custodial symmetry
\cite{custodial}, and if it were possible for the
underlying theory to respect this symmetry $\tilde{b}$ would be zero.
There must, however, be sufficient custodial violation to give rise to
the top-quark Yukawa coupling. In the absence of custodial symmetry,
dimensional analysis predicts a top Yukawa coupling of order $\kappa$.
The violation of custodial symmetry, therefore, introduces the small
parameter $y_t/\kappa \simeq 1/\kappa$ \cite{Chivukula:1996sn}. The
second term in eqn. (\ref{operators}) is then suppressed by this
amount\footnote{In principle, since the top-quark mass operator
  ``transforms'' as a custodial isospin $I=1$ operator and the second operator in
  eqn. (\protect\ref{operators}) as $I=2$, the suppression could be as
  much as $(y_t/\kappa)^2 \simeq 1/\kappa^2$. In practice, by varying
  $b\kappa$ between 1 and $4\pi$ in eqn. (\ref{cusoperators}) we expect
  (\ref{corrections}) will yield a reasonable estimate of the size
  of these effects.} and in a  custodially-symmetric
strongly-interacting underlying theory we expect the low
energy operators
\beq
-\, {a \over 2!\, \Lambda^2} \left\{ [D_\mu,D_\nu] \phi \right\}^\dagger[D^\mu,D^\nu]
\phi
+{b\, \kappa \over 2!\, \Lambda^2} (\phi^\dagger \stackrel{\leftrightarrow}{D^\mu} \phi) 
(\phi^\dagger \stackrel{\leftrightarrow}{D_\mu} \phi)~,
\label{cusoperators}
\eeq
where $a$ and $b$ are both ${\cal O}(1)$. These operators give rise to
the corrections
\beq
\Delta S = {4 \pi a  v^2\over  \Lambda^2},\hspace{1cm}
\& \hspace{1cm} \Delta T = {b\kappa v^2 \over \alpha\Lambda^2}
\label{corrections}
\eeq
where $\alpha$ is the electromagnetic coupling.

\begin{figure}
\epsfxsize 10cm \centerline
{\epsffile{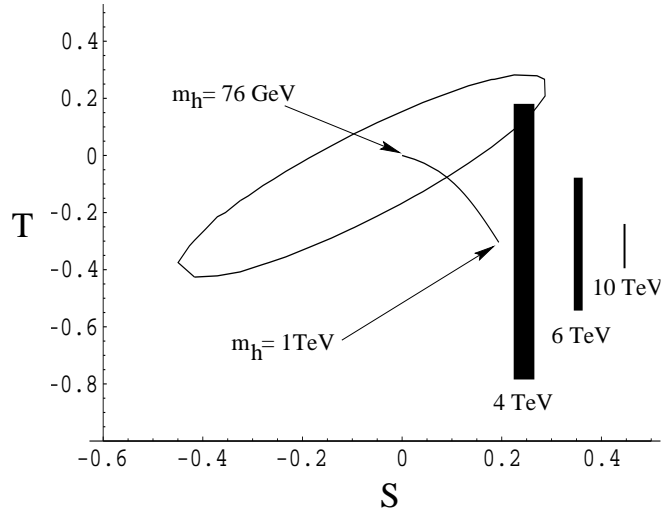}}
\caption{
  The oval demarks the area of the S-T plane compatible with the
  precision electroweak measurements at the $95 \%$ confidence level.
  The line is the trajectory of Higgs mass in the standard model from 76
  GeV to 1 TeV. The black rectangles show the natural size of
  corrections from the underlying physics for different scales $\Lambda$
  and varying $a$ and $b \kappa $ between $\pm 1$ (they should be
  centered on the point on the Higgs line corresponding to the Higgs
  mass of interest).}
\label{Fig2}
\end{figure}

In figure \ref{Fig2}, we display a fit to current electroweak data
\cite{LEPEWWG} from $Z^0$ measurements at LEP and SLD and the $W$ mass
measurements from LEP II and the Tevatron, allowing for the presence of
flavor-universal ``oblique'' contributions to the gauge-boson
self-energies
\cite{Peskin,others}.
The fit was done by incorporating the Peskin-Takeuchi $S$ and $T$
parameters \cite{Peskin}, defined using a reference
Higgs boson mass of 76 GeV, into the prediction of the electroweak
observables \cite{Burgess:1994vc}. The 95\% confidence contour in this
plane encloses the point $S=T=0$, displaying the agreement of the data
with the standard model with a light Higgs boson.  Changing Higgs mass
can be viewed as parametrically changing $S$ and $T$. The $S$ and $T$ 
dependence on the higgs mass has been calculated in \cite{Hagiwara:1997zm}. 
The curve giving the standard model prediction
varying the Higgs mass from 76 GeV to 1 TeV is also shown in figure
\ref{Fig2}. These considerations yield, at 95\% confidence level, a
limit of 230 GeV on the Higgs mass (120 GeV at $67\%$) which agrees well
with the fit of ref. \cite{Karlen}. The dependence of $S$ and $T$
on the higgs mass is only logarithmic and the bounds are thus very sensitive 
to the data (for example removing the SLD measurement of the left right
asymmetry increases the $95\%$ bound to of order 400 GeV \cite{Karlen}). The bounds
are also very sensitive to the inclusion of new
physics \cite{newphysics} as well.

In figure \ref{Fig2} we also show the natural size of the corrections
from the underlying physics (the operators in eqn. (\ref{operators})) as
error boxes (which should be centered on the appropriate point on the
Higgs mass curve to see the effect of the corrections for a given Higgs
mass) for different scales $\Lambda$ and varying $a$ and $b\kappa$
between $\pm 1$. Note that the error boxes are rather narrow -- it is
the shifts in $T$ induced by the underlying physics that are the most
relevant. For $\Lambda$ of order 5 TeV to 10 TeV, which are the upper
bounds on $\Lambda$ for Higgs bosons with masses of order 500 to 600
GeV, corrections to the $T$ parameter are of sufficient size that they
are not negligible in the context of the Higgs mass bounds.  If we allow
larger values of $a$ and $b \kappa$, then lower scales become disfavored
because the natural contribution to $T$ is too large (this is the
familiar problem found in accommodating the top mass in technicolor type
models \cite{isospin}) but the corrections from the underlying physics
remain important up to a scale of order 50 TeV! To reconcile a heavy
higgs with the precision data requires positive contributions to the $T$
parameter.  In fact, generically, isospin breaking does give rise to
positive contributions to $T$ ({\it e.g.} mass splitting in heavy
electroweak doublets, mixing with additional $U(1)$ gauge bosons etc)
and negative contributions are much harder to achieve. Alternatively,
negative contributions to $S$ could reconcile a heavy higgs but the
natural size of corrections from the underlying physics does not seem
compatible with this choice.

If we include the corrections (\ref{operators}) we may derive 
bounds in the higgs mass - $\Lambda$ plane. These are shown in figure 1 for
varying values of $b \kappa$ (we set $a = - b \kappa$ to provide
the most conservative bound but the dependence on the $S$ parameter is
small). For Higgs masses above 500 GeV the triviality bound itself
is in fact a stronger bound than the precision data including 
corrections from the underlying physics.   

\begin{figure}
\epsfxsize 15cm \centerline
{\epsffile{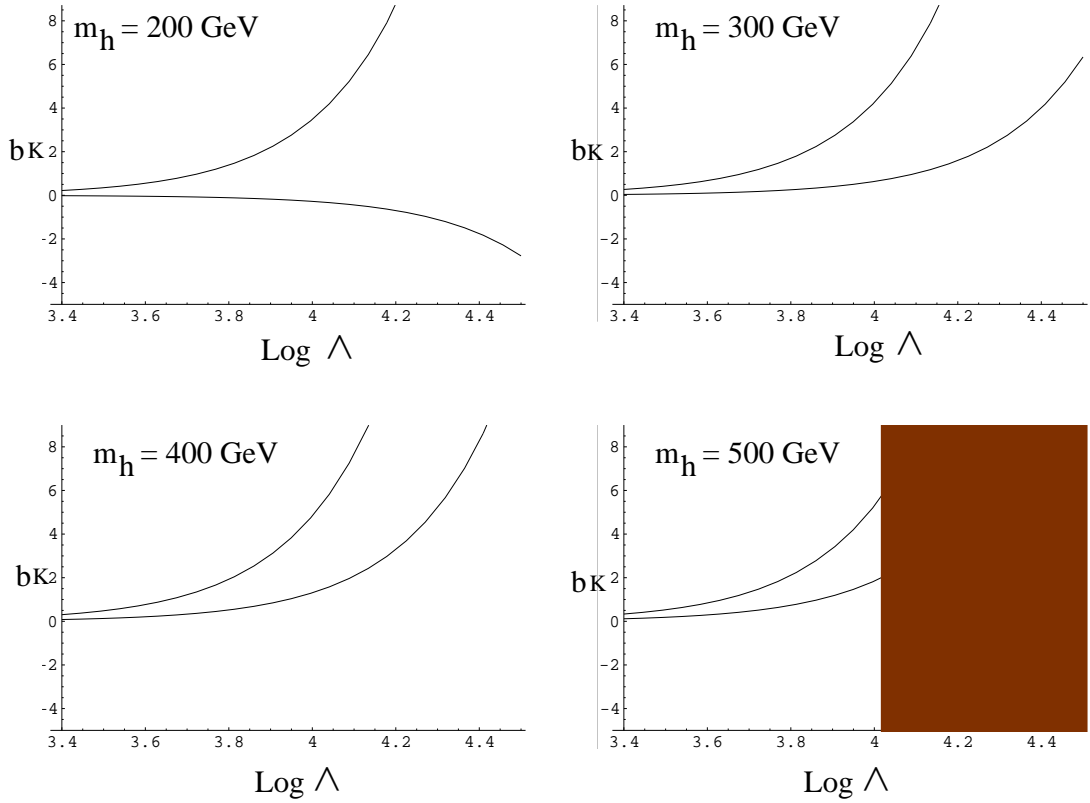}}
\caption{
  The value of $b \kappa$ compatible with the precision bounds as a
  function of $\Lambda$ for different values of $m_h$. The blocked off
  area in the last plot is forbidden by triviality. The values of cut
  off on the x axis varies logarithmically between 2500 -- 25000 GeV.  }
\label{Fig3}
\end{figure}

One may worry that $b\kappa$ must be very finely adjusted to make
heavier Higgs bosons consistent with precision electroweak measurements,
and that this situation is therefore not generic.  In figure \ref{Fig3}
we plot, for various Higgs boson masses, the values of $b\kappa$ which
are allowed by precision electroweak measurements as a function of
$\Lambda$. We see that, for Higgs masses up to 500 GeV and scales
$\Lambda$ of order 10 TeV or less, no unnatural adjustment of parameters
is required. 

That $\Lambda$ is naturally of order 10 TeV or higher, with smaller cut
offs requiring a greater degree of fine tuning of $b \kappa$,
corresponds to the result of ref. \cite{Chivukula:1996sn}. In
\cite{Chivukula:1997iw} it was argued that if the fundamental physics at
the scale $\Lambda$ does not respect flavor symmetries then neutral
meson mixing gives a constraint on the scale $\Lambda$ of order 20 TeV.
Higgs masses above 460 GeV would then be ruled out by triviality alone
and to reconcile higgs mass between 230-460 GeV would require larger
values of $b \kappa$.  This stronger constraint is more speculative than
that from the $T$ parameter because, whilst isospin is known to be
broken by the top Yukawa, a GIM-type mechanism suppressing
flavor-changing neutral-currents could arise from the underlying
physics.

The triviality of the higgs sector of the standard model requires that
the standard model with a heavy higgs is only an effective field theory
that must break down at relatively low scales of ${\cal O}(10\,{\rm
  TeV})$ or less. Even if the higgs is lighter, a low cut off may exist in
nature. Higher dimension operators in the effective higgs theory
suppressed by the cut off scale will contribute to the $S$ and $T$
parameters. Using dimensional analysis, we have estimated the expected
size of these corrections in a custodially-symmetric
strongly-interacting underlying theory.  Taking these operators'
coefficients to be of natural size gives sufficiently large
contributions to the T parameter to reconcile Higgs masses as large as
400-500 GeV with the precision data.

\bigskip


\centerline{\bf Acknowledgments} We thank Bogdan Dobrescu for comments
on the manuscript.

{\em This work was supported in part by the Department of Energy under
  grant DE-FG02-91ER40676.  NE is grateful for the support of a PPARC
  Advanced Research Fellowship.}


\newpage


\end{document}